\documentclass{physeauth}
\usepackage{graphicx}
\usepackage[intlimits]{amsmath}
\usepackage{amssymb}

\begin{document}

\begin{frontmatter}

\title{Interacting resonant level coupled to a Luttinger liquid:
Population vs. density of states}

%


\author{Moshe Goldstein}, 
\author{Yuval Weiss},
\author{Richard Berkovits}

\address{The Minerva Center, Department of Physics, Bar-Ilan University,
Ramat-Gan 52900, Israel}


\begin{abstract}
We consider the problem of a single level quantum dot coupled to the
edge of a one-dimensional Luttinger liquid wire by both a hopping term and
electron-electron interactions.
Using bosonization and Coulomb gas mapping of the Anderson-Yuval type we
show that thermodynamic properties of the level, in particular, its
occupation, depend on the various interactions in the system only through
a single quantity --- the corresponding Fermi edge singularity
exponent.
However, dynamical properties, such as the level density of states,
depend in a different way on each type of interaction.
Hence, we can construct different models, with and without interactions
in the wire, with equal Fermi
edge singularity exponents, which have identical 
population curves, although they originate from very different level
densities of states. The latter may either be regular or show a
power-law suppression or enhancement at the Fermi energy.
These predictions are verified to a high degree of accuracy using the
density matrix renormalization group algorithm to calculate the dot
occupation, and classical Monte Carlo simulations on the corresponding
Coulomb gas model to extract the level density of states.
\end{abstract}


\end{frontmatter}


\section{Introduction} \label{sec:intro}

The behavior of low-dimensional electronic systems has
been in the focus of many experimental and theoretical
studies in recent years. Such systems are important both because of
the fundamental interest in the strongly correlated physics they exhibit,
as well as their role as the building blocks for
creating nano-scale devices.
The one-dimensional case is a particularly interesting one.
When no symmetry is spontaneously broken and the one-dimensional
system is metallic,
its low energy dynamics is described by the Luttinger liquid
(LL) theory \cite{bosonization}.
The latter offers one of the clearest realizations
of non Fermi liquid physics. These systems have
been experimentally realized in a variety of ways, including narrow
quantum wires in semiconducting heterostructures, metallic nanowires,
and carbon nanotubes. A closely related concept is that of chiral LLs,
which describe the physics of edge states in the fractional quantum
Hall effect \cite{chang03}.
A natural question which arises is the effect of impurities on these
systems, either naturally occurring or artificially introduced (e.g.,
quantum dots and anti-dots).
While this topic has been investigated for some time,
most of the previous works are restricted to the study of transport
properties
\cite{bosonization,chang03,kane92,furusaki93,nazarov03},
while other phenomena have been only occasionally considered
\cite{furusaki02,sade05,weiss07,wachter07,weiss08,lerner08,bishara08,goldstein08}.

In this paper we investigate one of the simplest systems of this
kind, namely a single level attached to the end of a LL
(or, equivalently, a level in the vicinity of a chiral LL).
We include short-range interactions between the
charge of the level and the charges in its neighborhood in the wire.
Although transport properties are, of course,
not relevant here, many other phenomena can be investigated.
In this paper we compare thermodynamic properties, in particular,
the level population (which can be measured by placing a quantum point
contact in the vicinity of the level), 
to dynamic properties, such as the level density of states (LDoS; this
can be probed by tunnel spectroscopy).
In section \ref{sec:model_cg} we demonstrate, using the Anderson-Yuval
approach, that while thermodynamic properties are universal, and are
affected by the various interactions only through a single parameter
(identified as the Fermi edge singularity exponent), dynamic properties
are sensitive to the specific physics of the different interaction types.
Based on this analysis, in section \ref{sec:numerics}
we construct different systems, some LLs and some not,
which are tested numerically to have
the same dependence of the population on the level energy, although
their LDoSs (of which the population is an integral) are very different.
These results are discussed and summarized in section~\ref{sec:conclude}.

\section{Model and Coulomb gas analysis} \label{sec:model_cg}
The system described in the Introduction can be modeled by the
Hamiltonian $H=H_{\mathrm{w}}+H_{\mathrm{l}}+H_{\mathrm{lw}}$.
The first term is the wire Hamiltonian, described by the standard
Tomonaga-Luttinger model. In bosonized form, it is given by \cite{bosonization}:
\begin{equation}
\label{eqn:hw}
 H_{\mathrm{w}}= \frac{v}{2\pi}
 \int_{0}^{\infty}
 \left\{ \frac{1}{g} [\nabla \theta(x)]^2 + g [\nabla \phi(x)]^2  \right\}
 \mathrm{d}x,
\end{equation}
where the bosonic fields $\theta(x)$ and $\phi(x)$ obey the
commutation relation $[\theta(x),\phi(y)]=\mathrm{i}\pi\Theta(x-y)$, and
the boundary condition $\theta(0)=0$. Here $g$ and $v$ are the
usual interaction parameter and excitation velocity, respectively,
and $\Theta(x)$ is Heaviside's step function.
The second term of the full Hamiltonian describes the level:
$H_{\mathrm{l}}=\varepsilon_0 d^{\dagger}d$, where $d^{\dagger}$ ($d$)
is the level creation (annihilation) operator
and $\varepsilon_0$ is its energy.
Finally, the level and the wire are coupled by:
\begin{eqnarray}
\label{eqn:hlw}
H_{\mathrm{lw}} & = &
- \left[ \mathcal{T}_{\mathrm{lw}} d^{\dagger} \psi(0) +
\mathrm{H.c.} \right] + 
\nonumber \\ & &
\mathcal{U}_{\mathrm{lw}}
\left( d^{\dagger}d - \frac{1}{2} \right)
:\psi^{\dagger}(0) \psi(0):.
\end{eqnarray}
The first part of $H_{\mathrm{lw}}$ describes the level-wire hopping,
parametrized by a tunneling matrix element ${\mathcal{T}}_{\mathrm{lw}}$,
while the second part, in which the colons denote normal ordering, is a local level-wire
interaction of strength ${\mathcal{U}}_{\mathrm{lw}}$.
The electronic annihilation operator at the wire's edge is given by
$\psi(0)=\chi \mathrm{e}^{\mathrm{i}\phi(0)}/ \sqrt{2\pi a}$,
using Majorana Fermi operators $\chi$ and a short distance cutoff (e.g., a lattice spacing) $a$.

Following Anderson and Yuval's method \cite{yuval_anderson}, any quantity
of interest is expanded to all orders in ${\mathcal{T}}_{\mathrm{lw}}$. This
results in a series of correlation functions, which are calculated
for ${\mathcal{T}}_{\mathrm{lw}}=0$ \cite{si93,kamenev_gefen}. Because
of the level-wire interaction, there is a potential at the edge of the
wire which flips between $\pm {\mathcal{U}}_{\mathrm{lw}}$ when each hopping occurs,
i.e., we have a sequence of Fermi edge singularity events \cite{noziers69}.

Let us first discuss the thermodynamic properties of the model, e.g.:
the level population (to be denoted by $n_{\mathrm{level}}$),
entropy and the specific heat. These can be expressed
through the partition function $Z$ of the model and its derivative with
respect to the parameters of the system, such as
the level energy $\varepsilon_0$ and the temperature $T$.
In the Anderson-Yuval approach, the partition function
acquires the form of a grand canonical partition function of a
classical system of particels (hopping events) residing on
the imaginary time axis of the original quantum model
(i.e., on a circle whose circumference is the inverse of the
temperature $T$ of the level-wire system).
We assign to a particle corresponding to
hopping of an electron from the level to the wire a positive
charge, and to a particle describing the reverse process
a negative charge.
Thus, the charges must be alternating in sign,
and their total number has to be even.
Denoting the position of the $i$'th particle by $\tau_i$ and
the sign of the charge of the first one by $s$, we obtain:
\begin{multline} \label{eqn:cg1}
Z = \sum_{ \substack{N=0 \\ s=\pm1}}^{\infty}
\left( \frac{\Gamma_0 \xi_0}{\pi} \right)^N
\int_0^{1/T} \frac{d\tau_{2N}}{\xi_0}
\int_0^{\tau_{2N}-\xi_0} \frac{d\tau_{2N-1}}{\xi_0}
\dots \\
\int_0^{\tau_3-\xi_0} \frac{d\tau_2}{\xi_0}
\int_0^{\tau_2-\xi_0} \frac{d\tau_1}{\xi_0}
\exp \left[ - H_{\mathrm{CG}} (\{ \tau_i \}, s) \right] ,
\end{multline}
so that the particles have a fugacity $\sqrt{\Gamma_0\xi_0/\pi}$,
where $\xi_0$ is a short time (ultraviolet) cutoff and $\Gamma_0$ is the
level width (an expression for which is given below).
The Hamiltonian of the classical system reads:
\begin{multline} \label{eqn:cg2}
H_{\mathrm{CG}}( \{ \tau_i \}, s ) = \\
\alpha_{\mathrm{FES}} \sum_{i<j =1}^{2N} (-1)^{i+j}
\ln \left\{ \frac{ \pi T \xi_0 }
{\sin [ \pi T (\tau_j-\tau_i) ] } \right\} + \\
\frac{\varepsilon_0}{T} \left\{ \frac{1}{2} -
s \left[ T\sum_{i=1}^{2N} (-1)^i \tau_i - \frac{1}{2} \right] \right\}.
\end{multline}
The first part of this expression describes an interaction between
the particles, which is similar in form to the electrostatic interaction
between charged rods, giving Eqs.~(\ref{eqn:cg1}-\ref{eqn:cg2}) the
name ``Coulomb gas expansion''.
The (absolute value of the) charge of each particle is
$\sqrt{\alpha_{\mathrm{FES}}}$, the square root of the Fermi
edge singularity exponent of the model.
This quantity is defined through the behavior of the zero-temperature
Green function of $d^{\dagger} \psi(0)$ for $\mathcal{T}_{\mathrm{lw}}=0$,
which, for long time $\tau$ decays as
$\tau^{-\alpha_{\mathrm{FES}}}$.
We will give expressions for this quantity below.
For a wire of a finite length $L$ at zero temperature,
the Coulomb gas interaction is given by
$\ln \left\{ \frac{ \pi \xi_0 v / L }
{\sinh [ \pi v (\tau_j-\tau_i)/L ] } \right\}$.
At finite temperatures it takes the form of a logarithm of an elliptic
function \cite{bosonization}. 
The other part of the Coulomb gas Hamiltonian is simply equal
to the total length of the imaginary time intervals in
which the level is occupied, multiplied by $\varepsilon_0$, the
energy of an occupied level. It is analogous to an electric
field applied on the classical system of charges.

Up to now, the values of $\alpha_{\mathrm{FES}}$ and $\Gamma_0$ have
not been specified.
When there are no intra-wire interactions (i.e., $g=1$), we have the
usual interacting resonant level model, in which case
$\alpha_{\mathrm{FES}} = \left( 1 - \frac{2}{\pi} \delta \right)^2$, and
$\Gamma_0 = \pi \left| {\mathcal{T}}_{\mathrm{lw}} \right|^2 \rho_0 \cos(\delta)$,
where $\delta=\tan^{-1} (\pi \rho_0 {\mathcal{U}}_{\mathrm{lw}}/2)$ is the phase
shift experienced by electrons at the Fermi surface in the wire due to the
level-wire interaction, and $\rho_0$ is the local density of states
in the wire edge at the Fermi energy \cite{noziers69,fabrizio95}.
For an interacting wire ($g \neq 1$) the situation is somewhat more complicated.
Since there is no backscattering in this problem, calculations using bosonization
\cite{bosonization} yield
$\alpha_{\mathrm{FES}}=(1-g{\mathcal{U}}_{\mathrm{lw}}/\pi v_s)^2/g$,
and $\Gamma_0=\pi \left| {\mathcal{T}}_{\mathrm{lw}} \right|^2 \rho_0$.
Comparing these expressions with the previous ones
in the limit of $g=1$ (noninteracting wire),
we see that the results of bosonization
replace the phase shift $\delta$ by its first Born approximation value.
Thus, the values of $\alpha_{\mathrm{FES}}$ and $\Gamma_0$ in any particular
systems are renormalized by irrelevant terms not appearing in the
Tomonaga-Luttinger Hamiltonian~(\ref{eqn:hw}).
It is natural to expect that, taking these effects into account,
for a general model we would get
$\alpha_{\mathrm{FES}} =
\frac{1}{g} \left( 1 - \frac{2g}{\pi}\delta_{\mathrm{eff}} \right)^2$,
for some effective phase shift $\delta_{\mathrm{eff}} \in [-\pi/2,\pi/2]$,
so that $\Gamma_0$ will be given by
$\pi \left| {\mathcal{T}}_{\mathrm{lw}} \right|^2 \nu_0 \cos(\delta_{\mathrm{eff}})$.
This can be shown to be valid for models in which $\alpha_{\mathrm{FES}}$
can be exactly evaluated \cite{goldstein08}.
Finally, it should be noted that in general $\varepsilon_0$ is also
modified by a term representing the difference in the total energy of the wire
caused by the potential applied on it by an empty or a filled level.
This correction can also be related to the phase shifts.
Its value, however, vanishes at half filling due to particle-hole symmetry,
which is the case in our numerical calculations.

Let us now turn to the LDoS, which we shall denote by
$\rho_{\mathrm{level}}(\omega)$. This quantity is equal
(up to a factor of $-1/\pi$) to the
imaginary part of the level retarded Green function, which in turn
can be obtained by analytic continuation from the corresponding Matsubara Green
function in the upper complex frequency plane \cite{mahan}.
In the imaginary time domain the Matsubara Green function is defined by
$G_{\mathrm{l}}(\tau) \equiv
-\mathrm{Tr}\{{\hat{T}}_{\tau} e^{-H/T} a(\tau) a^\dagger(0)\}/Z$,
where ${\hat{T}}_{\tau}$ is the imaginary time ordering operator. 
The numerator of this expression can be given a Coulomb-gas
representation, which has the same form as
Eqs.~(\ref{eqn:cg1}-\ref{eqn:cg2}) with two additional charges
of sizes $\pm (\sqrt{\alpha_{\mathrm{FES}}}-1/\sqrt{g})$, inserted at
$\tau$ and at the origin, respectively.
These charges correspond to the level creation and annihilation
operators appearing in the definition of the Green function.

From the above results we can immediately see an interesting distinction
between the Coulomb-gas expression for the partition function and that
for the LDoS:
The former contains only three parameters: $\Gamma_0$, $\varepsilon_0$, and
$\alpha_{\mathrm{FES}}$, while the latter explicitly depends on $g$ too.
As both the interactions in the wire and the level-wire interactions affect
the partition function mainly though a single combination ---
the Fermi edge singularity exponent $\alpha_{\mathrm{FES}}$, thermodynamics
cannot be used to distinguish between the different interaction types.
In other words,
one can construct very different models, whose interactions differ in
strength and even in sign, which will have the same thermodynamic properties,
provided $\Gamma_0$, $\varepsilon_0$, and $\alpha_{\mathrm{FES}}$
are indeed the same.
On the other hand the LDoS, which depends explicitly 
on $g$, can be used to disentangle the effects of intra-wire and level-wire
interaction, and will thus behave differently for these different systems.
In fact, it can be expected that at low energies
the LDoS should behave like the tunneling density of states near the
end of a LL, i.e., it will vary as $[\max ( |\omega|, T, v/L)]^{1/g-1}$
\cite{bosonization}. Thus, it exhibits a power law,
which depends only on the LL parameter $g$ and not
on $\alpha_{\mathrm{FES}}$, i.e., on the interactions in the wire but
not on the level-wire coupling.
These considerations thus give rise to a quite surprising possibility:
In spite of the fact that the level population is the integral of
the LDoS times the Fermi-Dirac distribution function,
a LL system (i.e., a system with $g \neq 1$) may have the same dependence
of the level occupancy on $\varepsilon_0$ as a Fermi liquid
system (for which $g=1$) with a properly chosen level-wire coupling, although the LDoS
is obviously different in the two cases. In the next section we will
demonstrate this to hold quantitatively, using numerical calculations.

\begin{figure}
\includegraphics[width=8cm,height=!]{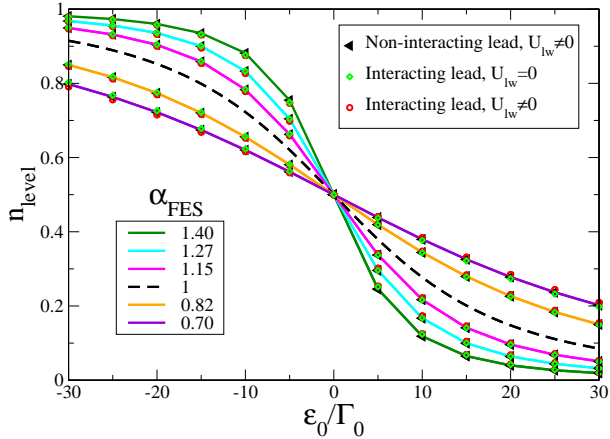}
\caption{\label{fig:nlevel}
(Color online)
DMRG results for the level population as a function of its energy,
for three different models denoted by the three different symbol types.
The curves on which the symbols reside (which serve as a guide to the eye)
correspond to the various $\alpha_{\mathrm{FES}}$ values
(the larger $\alpha_{\mathrm{FES}}$ the narrower the curve and vice versa).
See the text for further details.
}
\end{figure}

\begin{figure}
\includegraphics[width=8cm,height=!]{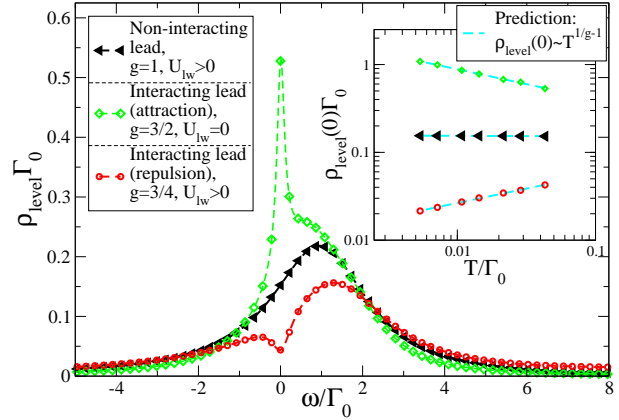}
\caption{\label{fig:ldos}
(Color online)
Monte Carlo results for the LDoS as a function of frequency, measured from the
Fermi energy for three different models, all with $\alpha_{\mathrm{FES}}=2/3$,
$\epsilon_0/\Gamma_0=0.7$, and $T/\Gamma_0=0.04$.
The three curves correspond to three different values of $g$, as indicated in
the legend.
Inset: temperature dependence of the LDoS at the Fermi energy (symbols),
together with the expected power-low behavior (line).
See the text for further details.
}
\end{figure}

\section{Numerical results} \label{sec:numerics}

In this section we show numerical data confirming the results of the previous
section, i.e., that the level population is universal (equal for
different models), although the LDoS is not.

To find the level occupation we have performed density matrix renormalization
group (DMRG) \cite{white93} calculations on a particular realization of a LL
wire represented by a half-filled $N$-site tight binding chain with nearest
neighbor interactions. In this case, the wire and level-wire Hamiltonians are
given by:
\begin{eqnarray}
 H_{\mathrm{w}} & = &
   \sum_{i=1}^{N-1}
   \left ( -t c_{i}^{\dagger} c_{i+1} +  \mathrm{H.c.} \right )
   +
   \nonumber \\ & &
   U \left( c_{i}^{\dagger} c_{i} -\frac{1}{2} \right)
     \left( c_{i+1}^{\dagger} c_{i+1} -\frac{1}{2} \right), \\
 H_{\mathrm{lw}} & = &
   - \left( t_{\mathrm{lw}} c_1^{\dagger} d + \mathrm{H.c.} \right)
   +
   \nonumber \\ & &
   U_{\mathrm{lw}}
     \left( d^{\dagger} d -\frac{1}{2} \right)
     \left( c_{1}^{\dagger} c_{1} -\frac{1}{2} \right),
\end{eqnarray}
where $c^{\dagger}_{i}$ ($c_i$) is a creation (annihilation) operator
for an electron at the wire's $i$'th site, $t$ and $U$ are the nearest-neighbor
hopping and interaction strengths along the chain, and $t_{\mathrm{lw}}$, $U_{\mathrm{lw}}$
denote the corresponding quantities in the level-wire coupling term.
The latter are related to the parameters of the continuum level-wire coupling
Hamiltonian (\ref{eqn:hlw}) by
${\mathcal{T}}_{\mathrm{lw}} = t_{\mathrm{lw}} \sqrt{a}$, and
${\mathcal{U}}_{\mathrm{lw}} = U_{\mathrm{lw}} a$,
$a$ being the lattice spacing.
Using boundary conformal field theory
arguments and the Bethe ansatz, it can be shown that for this model
\cite{goldstein08}:
\begin{equation} \label{eqn:alpha_bethe}
\alpha_{\mathrm{FES}} =
\frac{1}{g} \left[ 1 - \frac{2g}{\pi}
\tan^{-1} \left( \frac{ U_{\mathrm{lw}} }{ \sqrt{(2t)^2-U^2} }
\right) \right]^2,
\end{equation}
where $g=\pi/[2\cos^{-1}(-U/2t)]$.

In Fig.~\ref{fig:nlevel}
we show the level population as a function of its energy.
The different curves correspond to different
$\alpha_{\mathrm{FES}}$ values. On each
such curve there are symbols of three types, denoting DMRG
data in three different systems:
(i) $U=0$ but $U_{\mathrm{lw}} \neq 0$;
(ii) $U \neq 0$ but $U_{\mathrm{lw}} = 0$;
(iii) both $U \neq 0$ and $U_{\mathrm{lw}} \neq 0$.
The interactions in the three three models were chosen so as to give equal
$\alpha_{\mathrm{FES}}$ values, as denoted in the figure's legend
[For model (c) we used $U=\pm 0.5t$, with sign opposite to that of model (b)].
In all cases we have kept $\Gamma_0=10^{-4}t$ (by choosing $t_{\mathrm{lw}}$
appropriately), and $N=100v/t$. A block size of $256$ was used.
From the results one can clearly see that the population is indeed universal,
and is equal for different models provided $\alpha_{\mathrm{FES}}$
is the same, although the interaction strengths are
different in magnitude and sign for the three cases.

From the data of Fig.~\ref{fig:nlevel} we observe that the population
vs. level energy curves become wider as $\alpha_{\mathrm{FES}}$
becomes smaller and vice-versa.
Our previous results show that smaller $\alpha_{\mathrm{FES}}$
implies either $g>1$ (i.e., attractive interactions in the wire)
or $U_{\mathrm{lw}}>0$. This could be understood since
both options should enhance the effective level-wire hopping,
and thus lead to a wider population curve.
Indeed, for $g>1$ it is well known that the local
density of states at the edge of a LL (or at the middle
of a chiral LL) is enhanced at the Fermi energy \cite{bosonization}, so
level-wire hopping becomes effectively stronger.
Similarly, $U_{\mathrm{lw}}>0$ also facilitates larger tunneling
by the Mahan exciton effect \cite{noziers69}: 
due to the level-wire repulsion, when the level is empty the site at
the wire's end tends to be occupied and vice versa;
hence, it helps the electrons overcome the limitations on
hopping induced by the Pauli principle.
In the same way we can understand why larger $\alpha_{\mathrm{FES}}$
will cause narrower population curves.

To find the LDoS we used classical Monte-Carlo calculations
to compute the imaginary-time Green function from its Coulomb gas
representation. The results were then Fourier transformed into Matsubara
frequency domain, and analytically continued to obtain the retarded
Green function using the Pad\'{e} approximant technique \cite{pade}.
The results are shown in Fig.~\ref{fig:ldos}. The three curves have
parameters which are approximately equal to those of the three
models represented by the widest curve of Fig.~\ref{fig:nlevel}.
We immediately see that, although the populations are, to a very high
accuracy, equal in the three models, since $\alpha_{\mathrm{FES}}=2/3$
for all of them (actually, the Coulomb gas representation would
predict exactly identical occupations),
the LDoS are markedly different, having a maximum, a minimum, or no
special feature near the Fermi energy for $g>1$, $g<1$, and $g=1$, respectively.
In the inset we demonstrate that the LDoS at the Fermi energy has a power-law
dependence on temperature for all three cases, with the expected power of $1/g-1$.

\section{Conclusions} \label{sec:conclude}

To conclude, we have shown, both analytically and numerically,
that the thermodynamics of a level
coupled to the edge of a LL is universal, depending on only a few parameters.
Thus, thermodynamic quantities (in particular --- the level occupation)
of a level coupled to a LL can be equal to the corresponding quantities
in other systems which are Fermi-liquids.
This occurs although thermodynamic quantities are determined by dynamic
properties which are not universal.
In this work we have concentrated on the level occupation, which
is determined by the LDoS. The latter
exhibits a power law behavior at low frequency with a power directly related
to LL parameters, which cannot be reproduced by a Fermi-liquid.
Yet, this LL-specific power law does not reflect itself in a LL-specific
$n_{\mathrm{level}} (\varepsilon_0)$ dependence.
Thus, thermodynamics of quantum impurities
may not help to expose LL physics. Nevertheless, our results imply that
interesting phenomena in such systems may be studied on
equivalent models with non-interacting wires,
which are much easier to approach, both analytically and numerically
(using, e.g., Wilson's numerical renormalization group \cite{nrg}).

\section*{Acknowledgments}

We would like to thank Y. Gefen and A. Schiller for many useful discussions.
M.G. is supported by the Adams Foundation of the Israel Academy
of Sciences and Humanities.
Financial support from the Israel Science Foundation (Grant 569/07) is
gratefully acknowledged.



\end{document}